\documentclass[aps,twocolumn,prd,floatfix,preprintnumbers,superscriptaddress,notitlepage,nofootinbib]{revtex4-2}
\pdfoutput=1
\usepackage{graphicx, color}

\usepackage[letterspace=-10]{microtype} 

\usepackage{bm, amsmath, amsfonts, amssymb}
\usepackage{multirow, tabularx, dcolumn}
\usepackage{mathtools} 
\usepackage{booktabs}
\usepackage{placeins}
\usepackage{soul} 
\usepackage{ulem}

\usepackage[utf8]{inputenc} 
\usepackage{hyperref}
\usepackage{orcidlink}

\usepackage{xcolor}
\definecolor{greena}{rgb}{0.0, 0.5, 0.0}

\hypersetup{colorlinks=true,linkcolor=greena,citecolor=greena,menucolor=black,urlcolor=greena,filecolor=greena}

\graphicspath{{./}}
\renewcommand\arraystretch{1.25}
\allowdisplaybreaks

\newcommand{\tud}{\affiliation{Institut für Kernphysik, Technische Universität Darmstadt, 64289 Darmstadt, Germany}}
\newcommand{\hdu}{\affiliation{School of Sciences, Hangzhou Dianzi University, Hangzhou 310018, China}}
\newcommand{\sxu}{\affiliation{College of Physics and Electronic Engineering, Shanxi University, Taiyuan 030006, China}}

\begin{document}

\title{Towards the LHCb pentaquark modes in the single-charm sector}

\author{Chao-Wei Shen\orcidlink{0000-0002-3529-4606}}\email{shencw@hdu.edu.cn}\hdu
\author{Yong-Hui Lin\orcidlink{0000-0001-8800-9437}}\email[Corresponding author: ]{yonghui.lin@tu-darmstadt.de}
\tud
\author{Hao-Jie Jing\orcidlink{0000-0002-3651-5722}}\email{jinghaojie@sxu.edu.cn}
\sxu

\begin{abstract}
The mass spectrum of $\Lambda_c$ baryon family is investigated within the $DN$-$D^*N$ coupled-channel framework using two phenomenological approaches for the low-energy $D^{(*)} N$ interactions: the heavy quark effective theory and the flavor-symmetry-constrained effective Lagrangian method. 
It is shown that the LHCb pentaquark states have direct analogs in the $\Lambda_c$ family. Specifically, $\Lambda_c(2765)$, $\Lambda_c(2910)$, and $\Lambda_c(2940)$ mirror the patterns of $P_{c\bar{c}}(4312)$, $P_{c\bar{c}}(4440)$, and $P_{c\bar{c}}(4457)$, respectively. 
These mirror pentaquark modes offer valuable insights into the quantum numbers of the two heavy $P_{c\bar{c}}$ states, which remain experimentally undetermined.
Further exploration of such correlations among exotic hadronic states across different flavor sectors will be crucial for developing a comprehensive theoretical understanding of the modern hadron spectrum.    
\end{abstract}

\maketitle

\newpage

\section{Introduction}\label{sec:intro}
Resolving the composition of matter is one of the most fundamental questions in the natural sciences. This quest has driven the development of modern particle physics, which is governed by the Standard Model, the most comprehensive and fundamental theory established to date. 
A milestone in this journey was the discovery of the eightfold way (also known as the quark model later), which achieved remarkable success in describing the observed hadron spectrum in the 1960s~\cite{Gell-Mann:1961omu,Gell-Mann:1964ewy,Zweig:1964ruk,Zweig:1964jf}. In this framework, hadrons are identified as the quark-antiquark mesons and the three-quark baryons. 
The quark model provided valuable insight into the underlying $SU(3)$ symmetry of the strong interaction, ultimately paving the way for the development of quantum chromodynamics (QCD), the most fundamental theory describing the strong force. 
The challenge emerged with the advent of the new century. Since 2003, when the first signal of a hadron with exotic nature was established through the observation of $X(3872)$ (also known as $X_{c1}(3872)$~\cite{ParticleDataGroup:2024cfk}) in exclusive $B^{\pm}\!\to\! K^\pm \pi^+\pi^- J/\psi$ decays by the Belle Collaboration~\cite{Belle:2003nnu}, a steadily growing family of exotic hadron candidates, lying beyond the conventional quark model of quark-antiquark mesons and three-quark baryons, has been reported experimentally and extensively investigated theoretically (see Refs.~\cite{Hosaka:2016pey,Esposito:2016noz,Guo:2017jvc,Olsen:2017bmm,Karliner:2017qhf,Kalashnikova:2018vkv,Brambilla:2019esw,Meng:2022ozq,Liu:2024uxn,Chen:2024eaq} for recent reviews).
Unraveling the internal structure of these exotic hadrons has become a central topic in hadron physics.   

In 2015, the LHCb Collaboration reported the first experimental evidence of pentaquark candidates, that is, $P_{c\bar{c}}(4380)$ and $P_{c\bar{c}}(4450)$ signals observed in $\Lambda_b^0\to J/\psi p K^-$ decays~\cite{LHCb:2015yax}. Four years later, with a data sample nine times larger, LHCb provided a more precise mass spectrum, identifying three narrow structures: $P_{c\bar{c}}(4312)$, $P_{c\bar{c}}(4440)$ and $P_{c\bar{c}}(4457)$~\cite{LHCb:2019kea}. 
Notably, the reported masses of these $P_{c\bar{c}}$ states are in remarkable agreement with earlier theoretical predictions obtained from the coupled-channel dynamical generation framework~\cite{Wu:2010jy,Wu:2010vk}. Motivated by their proximity to relevant thresholds, these LHCb pentaquark candidates have been naturally and successfully interpreted as hadronic molecular states composed of $\bar{D}^{(*)}\Sigma_c^{(*)}$ channels~\cite{Liu:2019tjn,Du:2019pij}.
Furthermore, the universality of near-threshold molecular formation in systems of open- and hidden-heavy-quark hadron pairs with attractive interactions has been highlighted in Ref.~\cite{Dong:2020hxe}, triggering an explosion of theoretical predictions for molecular candidates across various sectors~\cite{Dong:2021bvy,Dong:2021juy}. 
The LHCb pentaquark discoveries have firmly established the hadronic molecule as a compelling alternative to the conventional quark-antiquark mesons and three-quark baryons in describing exotic hadrons (see, e.g., Ref.~\cite{Guo:2017jvc} for a recent review).

Very recently, Ref.~\cite{Yue:2024paz} proposed an intriguing analogy between the $\Lambda_c$ baryon spectrum and the LHCb pentaquark states: 
$\Lambda_c(2910)$ and $\Lambda_c(2940)$, located near the $D^*p$ threshold,
resemble the $P_{c\bar{c}}(4440)$ and $P_{c\bar{c}}(4457)$ near the $\bar{D}^*\Sigma_c$ threshold.
To date, eight $\Lambda_c$ baryons have been observed according to the latest Review of Particle Physics~\cite{ParticleDataGroup:2024cfk}, ranging from the ground state $\Lambda_c(2287)$ to $\Lambda_c(2940)$, with $\Lambda_c(2765)$ and $\Lambda_c(2910)$ being less well established.
The charmed baryon spectrum has been extensively investigated in a variety of theoretical frameworks (see Refs.~\cite{Roberts:2007ni,Klempt:2009pi,Crede:2013kia,Li:2021iwf,Cheng:2021qpd,Luo:2025sns} for reviews).
Most observed states follow quark-model expectations~\cite{Capstick:1986ter,Ebert:2011kk}, notably forming two nearly parallel Regge trajectories, $\Lambda_c(2287)(1S)$-$\Lambda_c(2625)(1P)$-$\Lambda_c(2880)(1D)$ and $\Lambda_c(2595)(1P)$-$\Lambda_c(2860)(1D)$~\cite{Cheng:2021qpd}.
In contrast, the $\Lambda_c(2940)$, located just below the $D^*p$ threshold, deviates from this pattern and has been widely interpreted as an $S$-wave $D^*N$ molecular state~\cite{He:2006is,Dong:2010xv,He:2010zq,Luo:2019qkm,Wang:2020dhf,Yan:2023ttx,Yue:2024paz}.
The attractive interaction between charmed mesons and the nucleon has further inspired searches for possible three-body bounded heptaquark states in the $DD^*N$ system~\cite{Luo:2022cun,Montesinos:2024eoy}.

Motivated by the striking correspondence between the $\Lambda_c$ spectrum and the LHCb pentaquark states, we revisit the $\Lambda_c$ family within a $DN$-$D^*N$ coupled-channel framework, employing two complementary low-energy approaches for the $D^{(*)} N$ interactions: the heavy-quark effective theory and a flavor-symmetry--constrained effective Lagrangian method. 
We find that $\Lambda_c(2765)$, $\Lambda_c(2910)$, and $\Lambda_c(2940)$ can be well established as the $S$-wave $1/2^-$-$DN$, $1/2^-$-$D^*N$ and $3/2^-$-$D^*N$ molecules, respectively, within a heavy-quark--symmetry framework analogous to that of Refs.~\cite{Du:2019pij,Du:2021fmf}, where the $P_{c\bar{c}}(4312)$, $P_{c\bar{c}}(4440)$ and $P_{c\bar{c}}(4457)$ are successfully built from the coupled $\bar{D}\Sigma_c$-$D^*\Sigma_c$ dynamics. 
The flavor-symmetry--constrained effective Lagrangian method yields a consistent mass spectrum. 

\section{$DN$-$D^*N$ system in the heavy quark effective theory}\label{sec:HQET}
To obtain a rough estimate of the spectrum of the coupled $DN$-$D^*N$ systems, we construct a short-range effective field theory to describe the interactions between charm mesons and the nucleon, where only the contact terms are retained at leading order (LO). 
Following the treatment of the $\bar{D}^{(*)}\Sigma_c^{(*)}$ systems~\cite{Liu:2019tjn,Du:2019pij}, the number of independent low-energy constants (LECs) can be reduced by imposing heavy-quark symmetry constraints.
To this end, it is convenient to introduce the heavy-light spin basis, $|s_Q\otimes j_\ell\rangle$, where $s_Q$ denotes the total spin of the heavy quarks in the channel and $j_\ell$ represents the total angular momentum of the light degrees of freedom. 
Working in this the heavy-light spin basis, one can express the charm mesons and the nucleon as $\left|\frac12 \otimes \frac12\right\rangle$ and $\left|0 \otimes \frac12\right\rangle$ multiplets, respectively. 

Near the threshold, higher partial-wave contributions are suppressed due to the small relative momentum between threshold particles. 
Consequently, the $S$-wave short-range $D^{(*)}N$ potentials dominate the $DN$-$D^*N$ dynamics. 
The transformation between the $S$-wave $D^{(*)}N$ states and the corresponding 
heavy-light spin bases is given by
\begin{align}\label{eq: basis}
    \begin{pmatrix} |DN \rangle  \\ |D^*N \rangle  \end{pmatrix}_{1/2}&=\begin{pmatrix} \frac12 & \frac{\sqrt{3}}{2}  \\  \frac{\sqrt{3}}{2} & -\frac12 \end{pmatrix}\begin{pmatrix} \left|\frac12 \otimes 0 \right\rangle  \\ \left|\frac12 \otimes 1 \right\rangle  \end{pmatrix},\\
     |D^*N \rangle _{3/2}&=\left|\frac12 \otimes 1 \right\rangle,
\end{align}
where the subscripts on the left-hand side denote the total angular momentum of the channel. These decomposition relations show that, in the heavy quark limit, there are only two independent transitions between the $D^{(*)}N$ channels, which are given by
\begin{align}
    &C_0\equiv \left\langle s_Q\otimes 0\right| \hat{\cal H}_I \left| s_Q\otimes0\right\rangle,\notag\\
    &C_1\equiv \left\langle s_Q\otimes 1\right| \hat{\cal H}_I \left| s_Q\otimes1\right\rangle.
\end{align} 
Then the contact potentials for the $S$-wave $DN$-$D^*N$ coupled systems from the LO short-range
effective field theory can be written as
\begin{align}\label{eq: VLO}
    &V^{1/2^-}=\begin{pmatrix} \frac14C_0+\frac34 C_1 & \frac{\sqrt{3}}{4}C_0-\frac{\sqrt{3}}4C_1  \\  \frac{\sqrt{3}}{4}C_0-\frac{\sqrt{3}}4C_1 & \frac34C_0+\frac14 C_1 \end{pmatrix},\notag\\
    &V^{3/2^-}= C_1,
\end{align}
with $C_0$ and $C_1$ two dimensionless LECs.

For these constant potentials, the mass spectrum of the $DN$-$D^*N$ coupled systems can be readily obtained by solving the on-shell factorized Lippmann-Schwinger equation, yielding:
\begin{equation}\label{eq: Tmatrix}
    T=(1-VG)^{-1} V.
\end{equation}
The relativistic two-point function regularized within the dimensional regularization scheme is adopted for better numerical performance,
\begin{align}\label{eq: GDR}
    &G(\sqrt{s},\mu,m_i,m_j)=\frac1{16\pi^2}\bigg(a(\mu)+\log\frac{m_i^2}{\mu^2}+\frac{s+\Delta}{2s}\log\frac{m_j^2}{m_i^2}\notag\\
    &\phantom{}+\frac{\lambda^{1/2}(s)}{2s}\big[\log\left(s-\Delta+\lambda^{1/2}(s)\right)+\log\left(s+\Delta+\lambda^{1/2}(s)\right)\notag\\
    &\phantom{}-\log\left(-s+\Delta+\lambda^{1/2}(s)\right)-\log\left(-s-\Delta+\lambda^{1/2}(s)\right)\big]\bigg).
\end{align}
Here, $\Delta\equiv m_i^2-m_j^2$ and $\lambda(s)=s^2+m_i^4+m_j^4-2sm_i^2-2sm_j^2-2m_i^2m_j^2$ denotes the kinematic K\"{a}ll\'{e}n\ function. $a(\mu)$ is an unknown subtraction constant at the renormalization scale $\mu$, which is often taken as $1\,{\rm GeV}$. 
Conventionally, $a(\mu)$ is fixed by matching the above two-point function to
the one calculated using a three-momentum hard-cutoff scheme.
Our goal is to explore whether a solution for the LECs, $(C_0, C_1)$, exists that can reproduce the masses of $\Lambda_c(2765)$, $\Lambda_c(2910)$ and $\Lambda_c(2940)$ from the constructed $T$-matrix for the $S$-wave $DN$-$D^*N$ coupled-channel systems.

\begin{table}[htbp]
    \centering
    \caption{Two solutions for the LECs, $(C_0, C_1)$, when the subtraction constant $a=-2.1$ is fixed by matching the two-point function to that calculated in the three-momentum hard-cutoff scheme with $q_{\rm max}\approx1.0\,{\rm GeV}$.\label{Tab: solution_HQET}}
    \small
    \begin{tabular}{|c|c|c|c|c|c|}
        \hline
        \hline
        \multicolumn{3}{|c|}{$(C_0, C_1)$} & $(-145,-191)$ & $(-235,-153)$ & Exp.\\
        \hline
        State & RS  & $J^P$ & Pole [MeV] & Pole [MeV] & [MeV] \\
        \hline
        $\Lambda_c(2765)$ & $++$ & $\frac{1}{2}^-$ & $2763.6$ & $2765.4$ & $2766.6(24)$ \\
        $\Lambda_c(2910)$ & $+-$ & $\frac{1}{2}^-$ & $2939.6$ & $2907.0$ & $2914(7)$ \\
        $\Lambda_c(2940)$ & $+$  & $\frac{3}{2}^-$ & $2916.6$ & $2939.9$ & $2939.6(15)$ \\
        \hline
        \hline
    \end{tabular}
\end{table}
Two solutions that reproduce the $\Lambda_c$ mass spectrum are summarized in Tab.~\ref{Tab: solution_HQET}. 
Similar to the findings for the $\bar{D}^{(*)}\Sigma_c^{(*)}$ systems in Refs.~\cite{Liu:2019tjn,Du:2019pij}, these two solutions correspond to opposite spin-parity assignments for the higher $D^*N$ molecules: in one case, $\Lambda_c(2910)$ and $\Lambda_c(2940)$ are identified as the $1/2^-$ and $3/2^-$ $D^*N$ states, respectively, while in the other, the assignments are reversed.
In the RPP~\cite{ParticleDataGroup:2024cfk}, the $\Lambda_c(2940)$ is most likely assigned a spin-parity of $J^P=3/2^-$, as constrained by recent LHCb analyses~\cite{LHCb:2017jym}.
This close correspondence between the $\Lambda_c$ baryon family and the $P_{c\bar{c}}$ states thus supports assigning $J^P=3/2^-$ to the higher state in each corresponding doublet, implying a $3/2^-$ $\bar{D}^*\Sigma_c$ configuration for the $P_{c\bar{c}}(4457)$.   
Furthermore, the $\Lambda_c(2765)$, $\Lambda_c(2910)$ and $\Lambda_c(2940)$ mirror precisely the LHCb pentaquark modes in the single charm sector, providing compelling evidence that the universal near-threshold structures emergent in attractive open- and hidden-heavy-quark hadron pairs~\cite{Dong:2020hxe} may also manifest in systems with a single heavy quark.

\section{$DN$-$D^*N$ system in the effective Lagrangian method}\label{sec:efftive_lag}

Now let us turn to the dynamics of the $D^{(*)}N$ system within the effective Lagrangian framework, where the charm-nucleon interaction is modeled more realistically using one-boson-exchange potentials. 
The unitarized scattering $T$-matrix, constructed from these one-boson-exchange mechanisms, is obtained by solving the corresponding equations of motion, which are formulated as follows:
\begin{align}\label{eq: LSE}
    &T_{ij}(p^{\prime}, p, z) =  V_{ij}(p^{\prime}, p, z)  \nonumber \\
    &\phantom{xx} + \sum_k \int_0^{q_{\rm max}} \frac{q^2{\rm d}{ q}}{2\pi^2}  V_{ik}(p^{\prime}, q, z) G_k(q, z) T_{kj}(q, p, z),
\end{align}
Here $z$ denotes the center-of-mass energy, while $p$ ($p^{\prime}$) represents the magnitude of the center-of-mass three-momentum of the initial (final) state. The quantity $q$ is the three-momentum carried by the exchanged particle.
The subscripts $i$, $j$ and $k$ serve as channel indices, encoding information on spin, orbital angular momentum, total angular momentum, and isospin. 
To explore the hadronic molecular spectrum of the $D^{(*)}N$ systems, only the $S$-wave interaction potentials $V_{ij}(k,q,z)$ are considered. 
The two-body propagator $G_k(q, z)$ for the intermediate $k$-th channel is given by
\begin{align}
    G_k(q, z) = \frac1{4E_k(q)\omega_k(q)[z-E_k(q)-\omega_k(q)+i\varepsilon]},
\end{align}
where $E_k(q)$ and $\omega_k(q)$ are the on-shell energies of the baryon and meson, respectively.
Note that, as shown in Eq.~\eqref{eq: LSE}, a hard cutoff $q_{\rm max}$ is introduced to regularize the loop integral, ensuring that the low-energy dynamics dominate the possible formation of molecular states. In principle, $q_{\rm max}$ is a process-dependent unknown parameter and is varied within a typical energy range to estimate the theoretical uncertainty.

The interaction kernel $V$ iterated in Eq.~\eqref{eq: LSE} is constructed from the covariant effective Lagrangians~\cite{Wang:2022oof,Shen:2024nck}:
\begin{align}
    {\cal L}_{PPV} =& i g_{PPV} (P\partial^\mu P -\partial^\mu P P) V_\mu, \nonumber \\
    {\cal L}_{VVP} =& \frac{g_{VVP}}{m_V} \varepsilon_{\mu\nu\alpha\beta} \partial^\mu V^\nu \partial^\alpha V^\beta P, \nonumber \\
    {\cal L}_{VVV} =& i g_{VVV} \langle V^\mu [V^\nu, \partial_\mu V_\nu] \rangle, \nonumber \\
    {\cal L}_{BBP} =& \frac{g_{BBP}}{m_P} \bar{B} \gamma^\mu \gamma^5 \partial_\mu P B, \nonumber \\
    {\cal L}_{BBV} =& - g_{BBV} \bar{B} (\gamma^\mu - \frac{\kappa}{2m_B}\sigma^{\mu\nu}\partial_\nu) V_\mu B.
\end{align}
Here, $B$, $P$, and $V$ denote the baryon octet, the pseudoscalar mesons, and the vector mesons, respectively.
Both pseudoscalar- and vector-meson exchanges contribute to the $S$-wave $D^{(*)}N$-$D^{(*)}N$ transitions.
Approximate flavor symmetry~\cite{deSwart:1963pdg,Okubo:1975sc} is applied to relate the relevant coupling constants, while SU(4)-breaking effects are introduced to provide additional flexibility in the theoretical model. 
Specifically, we introduce two SU(4)-breaking parameters associated with the $g_{DD\rho}$ and $g_{D^*D^*\rho}$ vertices by defining $g^\prime_{DD\rho}=f_{D}g_{DD\rho}$ and $g^\prime_{D^*D^*\rho}=f_{D^*}g_{D^*D^*\rho}$.
The full expressions for the amplitudes, the SU(4) coupling relations, and other technical details are provided in Appendix~\ref{appendixA}.

The physical states dynamically generated by the $D^{(*)}N$ interactions appear as poles of the $T$-matrix obtained from Eq.~\eqref{eq: LSE}. 
For $J^P=1/2^-$, the $DN$ and $D^*N$ channels form a coupled system, yielding two poles: 
one below the $DN$ threshold, interpreted as a $DN$ bound state,
and another below the $D^*N$ threshold with a small imaginary part from the inelastic $DN$ coupling.
Inclusion of lighter decay channels (e.g., $\pi \Lambda_c$, $\eta \Lambda_c$, and $\pi \Lambda$)gives both poles finite widths.
For $J^P=3/2^-$, only the $D^*N$ channel contributes, producing a single pole on the real axis, corresponding to a $D^*N$ bound state.
Consequently, this framework predicts three $S$-wave states: a $DN$ bound state with $J^P = 1/2^-$, and two $D^*N$ bound states carrying $J^P = 1/2^-$ and $3/2^-$, respectively.

\begin{figure}[htbp]
    \centering
    \includegraphics[width=0.48\textwidth]{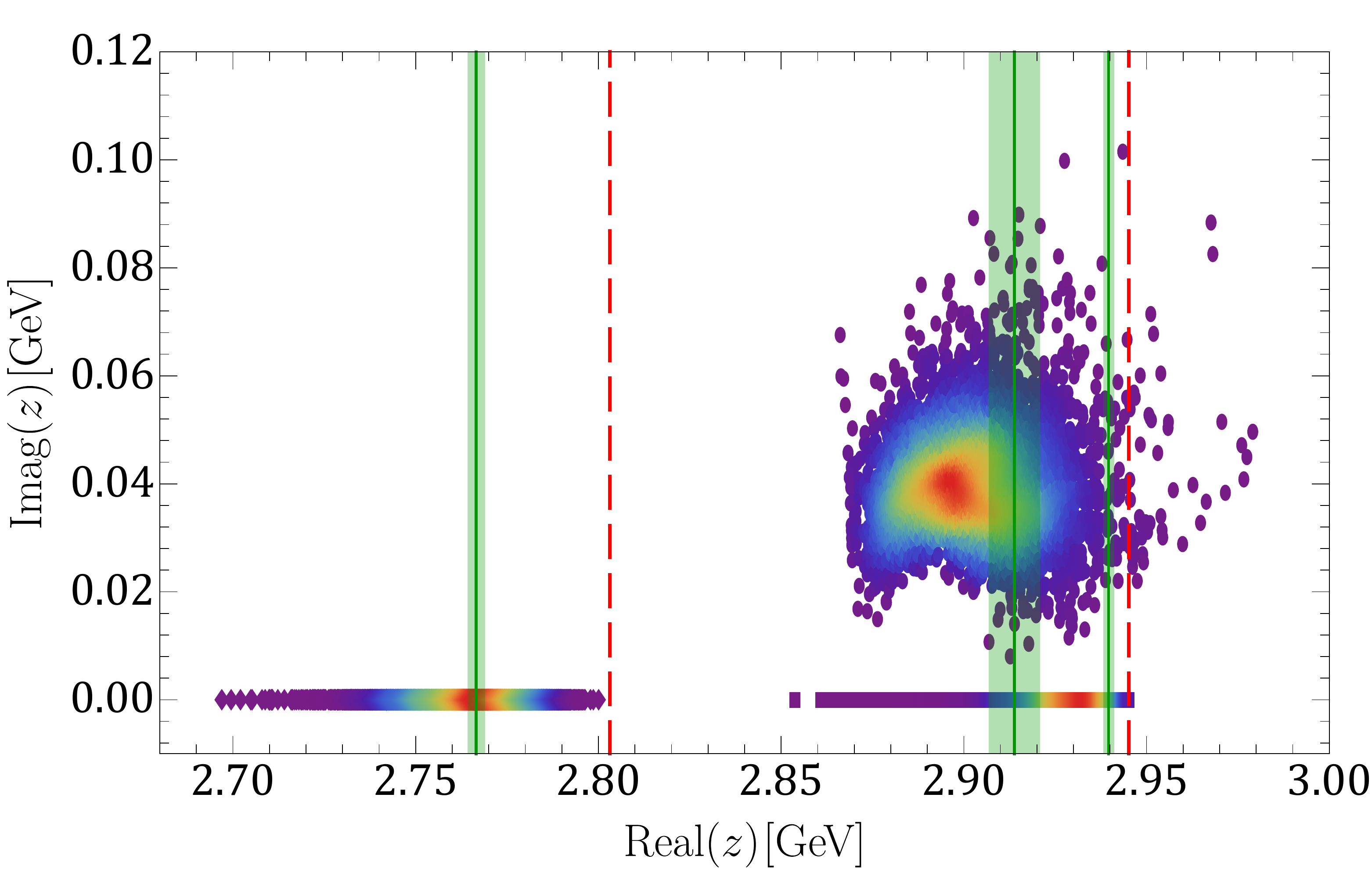}
    \caption{The distribution of pole positions obtained from 5000 different parameter samples. Colors indicate the density of the poles, with warmer colors representing higher densities. The two red dashed lines mark the $DN$ and $D^*N$ thresholds, while the green bars show the measured masses of $\Lambda_c(2765)$, $\Lambda_c(2910)$ and $\Lambda_c(2940)$ along with their experimental uncertainties.}
    \label{fig:pole_pos}
\end{figure}

To reproduce the masses of $\Lambda_c(2765)$, $\Lambda_c(2910)$, and $\Lambda_c(2940)$ with minimal free parameters, we found that it is sufficient to adjust only the SU(4)-breaking factors $f$ associated with $\rho$-exchange in the elastic channels, namely $f_D$ and $f_{D^*}$, together with a common three-momentum cutoff $q_{\rm max}$. 
Preliminary tests yielded central values $f_{D} = 1.3$, $f_{D^*} = 1.5$, and $q_{\rm max} = 800$~MeV.
Assigning a 10\% Gaussian uncertainty to each parameter, we generated 5000 Monte Carlo samples to estimate the theoretical uncertainty. 
The resulting distribution of the three eigenstates in the complex-energy plane is shown in Fig.~\ref{fig:pole_pos}.
Color gradients from red (highest density) through yellow and green to blue/violet (lowest density) visualize the pole populations, while green bars indicate the measured masses of $\Lambda_c(2765)$, $\Lambda_c(2910)$ and $\Lambda_c(2940)$ with experimental errors. 
The red-yellow cores of the pole clusters lie within these bars, demonstrating that the dynamically generated states can simultaneously reproduce the three observed $\Lambda_c$ masses.

Accordingly, $\Lambda_c(2765)$ is identified as a $1/2^-$ $DN$ bound state, while $\Lambda_c(2910)$ and $\Lambda_c(2940)$ correspond to $1/2^-$ and $3/2^-$ $D^*N$ states, respectively, consistent with the favored $3/2^-$ spin-parity of $\Lambda_c(2940)$ in the RPP~\cite{ParticleDataGroup:2024cfk}.
This conclusion agrees with the results obtained in the previous section within the heavy quark effective theory. The bound-state pattern in the single-charm sector can be directly compared with that in the hidden-charm sector.
Exploring correlations among exotic hadrons across flavor sectors, together with experimental confirmation from $pp$ scattering at HIAF or LHCb, $e^+e^-$ collisions at BESIII and STCF, is crucial for a comprehensive understanding of the hadron spectrum.

\section{Conclusion and outlook} \label{sec:summary}

In this paper, we have investigated $DN$-$D^*N$ coupled-channel interactions using both heavy quark effective theory and a flavor-symmetry-constrained effective Lagrangian approach.
We find that the mass spectrum of $S$-wave $D^{*}N$ systems, obtained from a short-range effective field theory with heavy-quark symmetry, can be matched to three exotic $\Lambda_c$ states: a $1/2^-$ $DN$ bound state and two $D^*N$ states with $J^P = 1/2^-$ and $3/2^-$, whose masses align with $\Lambda_c(2765)$, $\Lambda_c(2910)$, and $\Lambda_c(2940)$.
These assignments are further supported by a pole analysis based on a more realistic charm-nucleon interaction, modeling by the $t$-channel pseudoscalar and vector meson exchanges.
The resulting $D^{*}N$ molecules exhibit clear analogs to the hidden-charm pentaquarks observed by LHCb, with the $\Lambda_c$ spectrum mirroring the patterns of $P_{c\bar{c}}(4312)$, $P_{c\bar{c}}(4440)$, and $P_{c\bar{c}}(4457)$.  
Such mirror structures provide insight into the quantum numbers of experimentally undetermined $P_{c\bar{c}}$ and highlight the importance of exploring correlations among exotic hadrons across flavor sectors. 
Moreover, our results suggest that near-threshold structures observed in attractive open- and hidden-heavy-quark hadron pairs may also emerge in systems with a single heavy quark, pointing to a universal mechanism for hadronic molecule formation.

\section*{ACKNOWLEDGMENTS}

H.-J. J. is supported by the National Natural Science Foundation of China under Grants No.~12405100.

\appendix

\begin{appendix}

\section{Potentials and Couplings}\label{appendixA}
    
In total, five distinct exchange potentials ${\cal M}$ are involved in this work. 
Their explicit expressions, excluding the isospin factors, are given below:
\begin{align}
    &{\cal M}(D N \stackrel{\rho/\omega}{\rightarrow} D N) = g_a g_b \bar{u}_{3} (\gamma_\mu - \frac{\kappa}{2m_N}\sigma^{\mu\nu}iq_\nu) u_{1} \frac{(p_4+p_2)_\mu}{q^2-m^2}, \nonumber \\
    &{\cal M}(D N \stackrel{\pi/\eta}{\rightarrow} D^* N) = - i \frac{g_a g_b}{m_P} \bar{u}_{3} \gamma^\alpha \gamma^5 q_\alpha u_{1} \frac{(q-p_2)^\mu \epsilon^*_{4\mu}}{q^2-m^2}, \nonumber \\
    &{\cal M}(D N \stackrel{\rho/\omega}{\rightarrow} D^* N) = - \frac{g_a g_b}{m_V} \bar{u}_{3} (\gamma_\mu - \frac{\kappa}{2m_N}\sigma^{\mu\nu}iq_\nu) u_{1}  \nonumber \\
    &\phantom{xxxx} \times \frac{\varepsilon_{\alpha\beta\lambda\mu} p_4^\alpha \epsilon_4^{*\beta} q^\lambda}{q^2-m^2}, \nonumber \\
    &{\cal M}(D^* N \stackrel{\pi/\eta}{\rightarrow} D^* N) = - i \frac{g_a g_b}{m_P m_V} \bar{u}_{3} \gamma^\lambda \gamma^5 q_\lambda u_{1} \frac{\varepsilon_{\mu\nu\alpha\beta} p_2^\mu \epsilon_2^{\nu} p_4^\alpha \epsilon_4^{*\beta}}{q^2-m^2}, \nonumber \\
    &{\cal M}(D^* N \stackrel{\rho/\omega}{\rightarrow} D^* N) = - g_a g_b \bar{u}_{3} (\gamma_\mu - \frac{\kappa}{2m_N}\sigma^{\mu\nu}iq_\nu) u_{1}\frac{1}{q^2-m^2} \nonumber \\ 
    & \times \Big[ \epsilon_2^\alpha \epsilon_{4\mu}^{*} (q+p_4)_\alpha + \epsilon_{2\mu} \epsilon_4^{*\alpha} (p_2-q)_\alpha - \epsilon_2^\alpha \epsilon_{4\alpha}^* (p_4+p_2)_\mu \Big].
\end{align}
Herein, $q$ signifies the t-channel exchanged momentum.
Indices 1 and 3 refer to the incoming and outgoing baryons, respectively, while indices 2 and 4 refer to the incoming and outgoing mesons.

The isospin factors are obtained from
\begin{align}
  C_{IF}(I) =& \sum_{m_1,m_2,m_3,m_4} (m_1 m_2 I_z | I_1 I_2 I) (m_3 m_4 I_z | I_3 I_4 I) \nonumber \\
  &\times \langle I_3 m_3 I_4 m_4 | M^{iso} | I_1 m_1 I_2 m_2 \rangle,
\end{align}
where notation and conventions are given in Ref.\cite{Gasparyan:PhDthesis,Wang:2022oof}.
$(m s j |\ell S J)$ are the Clebsch-Gordan coefficients for the coupling of orbital angular momentum $\ell$ and spin $S$ to the total angular momentum $J$, with $m, s$ and $j$ the corresponding third components.
The isospin factors of different exchanged particles in $D^{(*)} N \to D^{(*)} N$ are given in Table~\ref{Tab: if}.
\begin{table}[htbp]
    \centering
    \renewcommand\arraystretch{1.3}
    \caption{The isospin factors for $D^{(*)} N \to D^{(*)} N$. \label{Tab: if}}
    \begin{tabular}{p{3.3cm}<{\centering}|p{1.8cm}<{\centering}|*{2}{p{1.2cm}<{\centering}}}
        \hline
        \hline
        Exchanged Particle & $C_{IF}\left(0\right)$ & $C_{IF}\left(1\right)$ \\
        \hline
        $\rho \ / \ \pi$ & -1 & 1 \\
        $\omega \ / \ \eta$ & 1 & 1 \\
        \hline
        \hline
    \end{tabular}
\end{table}

The final $S$-wave transition potential $V(p_4,p_2,z)$ used in Eq.~\eqref{eq: LSE} is then projected by means of
\begin{align}\label{eq:pwa}
    &V_{ij}(p_4,p_2,z)= 
    \frac{2\pi C_{IF}(0) { Y}_{0}^0(\hat{\mathrm{\mathbf{z}}})}{2J+1}\sum_{\substack{m_1 m_2, m_3m_4}}\int d\cos\theta\: { Y}_0^0(\cos\theta)^*
    \notag\\
    &\phantom{x}\times(m_1 m_2 m|S_1 S_2 S)(m_3 m_4 m|S_3 S_4 \bar{S})\: {\cal M}_{ij,m_1m_2}^{m_3m_4}(\vec{p}_4,\vec{p}_2,z),
\end{align}
where $\theta=\sphericalangle(\vec{p}_2,\vec{p}_4)$, and ${ Y}_l^m(x)$ are the spherical harmonic functions.
$S_1$ and $S_2$ denote the spins of the incoming and outgoing baryons, while $S_3$ and $S_4$ correspond to the incoming and outgoing mesons, with $m_i$ representing their respective third components. We use $J=S$ and $\bar J=\bar S$ to denote the quantum numbers of the $S$-wave initial and final $D^{(*)}N$ channels, respectively.

The coupling constants $g_{a,b}$ at all vertices are related through SU(4) flavor symmetry~\cite{Okubo:1975sc,Liu:2001ce,Dong:2009tg}. 
The corresponding relations for these couplings are given by:
\begin{align}
    g_{NN \pi}=& g_{BBP}, \
    g_{NN \eta}= \frac{\sqrt{3}}{5}g_{BBP}, \nonumber \\
    g_{NN \rho}=& g_{BBV}, \
    g_{NN \omega}= 3g_{BBV}, \nonumber \\
    g_{DD \rho}=& g_{PPV}, \ 
    g_{DD \omega}= g_{PPV}, \nonumber \\
    g_{D^*D \pi}=& g_{PPV}, \
    g_{D^*D \eta}= \frac1{\sqrt{3}}g_{PPV}, \nonumber \\
    g_{D^*D \rho}=& \frac1{\sqrt{2}}g_{VVP}, \ 
    g_{D^*D \omega}= \frac1{\sqrt{2}}g_{VVP}, \nonumber \\
    g_{D^*D^* \pi}=& \frac1{\sqrt{2}}g_{VVP}, \ 
    g_{D^*D^* \eta}= \frac1{\sqrt{6}}g_{VVP}, \nonumber \\
    g_{D^*D^* \rho}=& \frac1{\sqrt{2}}g_{VVV}, \ 
    g_{D^*D^* \omega}= \frac1{\sqrt{2}}g_{VVV},
\end{align}
where $g_{BBP}=0.989$, $g_{BBV}=3.25$, $\kappa_\rho=6.1$, $\kappa_\omega=0$, $g_{PPV}=4.27$, $g_{VVP}=-7.07$ and $g_{VVV}=2.298$~\cite{Janssen:1996kx,Ronchen:2012eg}.

\end{appendix}

\bibliography{refs}

\end{document}